\documentclass{iopart}             
\usepackage[T1]{fontenc}
\usepackage{graphicx}
\usepackage{url,graphics,epsfig}
\usepackage{amssymb}

\begin{document}
\jl{2}
%
%
%
\def\etal{{\it et al~}}
\def\newblock{\hskip .11em plus .33em minus .07em}
%
%
%
%
%
\setlength{\arraycolsep}{2.5pt}             

\title[Photoionization of  W] {Photoionization of the valence shells of the neutral tungsten atom}

\author{C P Ballance$^1$ and B M McLaughlin$^{2,3}\footnote[1]{Corresponding authors, 
					E-mail: ballance@physics.auburn.edu,b.mclaughlin@qub.ac.uk}$}

\address{$^1$Department of Physics,  206 Allison Laboratory,
                            Auburn University, Auburn, AL 36849, USA}

\address{$^2$Centre for Theoretical Atomic, Molecular and Optical Physics (CTAMOP),
                          School of Mathematics and Physics, The David Bates Building, 7 College Park,
                          Queen's University Belfast, Belfast BT7 1NN, UK}

\address{$^3$Institute for Theoretical Atomic and Molecular Physics,
                          Harvard Smithsonian Center for Astrophysics, MS-14,
                          Cambridge, MA 02138, USA}

%
%

\begin{abstract}
Results from large-scale theoretical cross section calculations for the total photoionization of the 4f, 5s, 5p and 6s orbitals
 of the neutral tungsten atom using the Dirac Coulomb R-matrix approximation (DARC: Dirac-Atomic R-matrix codes) are presented.
Comparisons are made with previous theoretical methods and prior experimental measurements. In previous experiments 
 a time-resolved dual laser approach  was employed for the photo-absorption of metal vapours
 and photo-absorption measurements  on tungsten in a solid, using synchrotron radiation.

 The lowest ground state level  of neutral tungsten is $\rm 5p^6 5d^4 6s^2 \; {^5}D_{\it J}$, with $\it J$=0,
 and requires only a single dipole matrix for photoionization.
 To make a meaningful comparison  with existing experimental measurements, 
 we statistically average the large-scale
theoretical PI cross sections from the levels associated with the
 ground state $\rm 5p^6 5d^4 6s^2 \; {^5}D_{\it J}[{\it J}=0,1,2,3,4]$ levels 
 and the $\rm 5d^56s \; ^7S_3$ excited metastable level.
 As the experiments have a self-evident metastable component in their ground state measurement,
 averaging over the initial levels allows for a more consistent and realistic comparison to be made.  
 
 In the wider context, the absence of many detailed electron-impact 
 excitation (EIE) experiments for tungsten and its multi-charged ion stages 
 allows current photoionization measurements and theory to provide a road-map for future 
 electron-impact excitation, ionization and di-electronic cross section calculations 
 by identifying the dominant resonance structure and features  across an 
 energy range of  hundreds of eV. 
\end{abstract}
%
%
\pacs{32.80.Fb, 31.15.Ar, 32.80.Hd, and 32.70.-n}

\vspace{0.5cm}
\begin{flushleft}
Short title: Valence shell photoionization of neutral Tungsten\\
\vspace{0.5cm}
\submitto{\jpb: \today}
\end{flushleft}

\maketitle
%
%
%
\section{Introduction}

Material choice for the plasma facing components in fusion experiments is determined by competing 
desirables: on the one hand the material should have a high thermal conductivity, high threshold for melting 
and sputtering,  low erosion rate under plasma contact, and on the other hand as a plasma impurity it 
should not cause excessive radiative energy loss. 
 
For the ITER tokamak, currently under construction at Cadarache, France, Tungsten 
(symbol W, atomic number 74) is the favoured material for the wall regions of highest particle and 
heat load in a fusion reactor vessel, with beryllium for regions of lower heat and particle load. 
ITER is to start operation with a W-Be or W wall for the main D-D and D-T experimental programme. 
In support of ITER and looking ahead to  the prospect of a fusion reactor other experimental plasma 
groups are also considering tungsten, including  the ASDEX-Upgrade tokamak which now operates with 
an all-W wall and at JET, where a full ITER-like mixed  W-Be-C wall is being installed. 
Smaller-scale experiments involving tungsten tiles are carried out on other tokamaks. 
The aforementioned attractive properties of tungsten mentioned above must be 
weighed against the negative fact that 74 ion stages have a great capacity to radiate power away.
The consequence is that the burn condition for tungsten is not achievable for concentrations 
above 2$\times$10$^{-4}$, and therefore theoretical models to
accurately characterize impurity influx are required \cite{ITER}.   

Atomic processes are central to energy transfer in magnetically confined plasmas.  
The energy balance in fusion devices such as tokamaks depends critically on how the 
plasma interacts with the walls of the vessel, therefore demanding accurate cross 
sections and associated rates for a wide variety of collisional processes.  
 
These rates will enable us to understand and mitigate the causes of critical radiation
losses that in minuscule concentrations prevent ignition.
For modelling the behaviour of tungsten in a plasma a comprehensive
understanding of various collisional processes is required for many ion stages. Beyond the first 
few charge states, open 4d and 4f shell configurations require 
complicated atomic structure calculations, and equally 
demanding electron-impact collisional calculations \cite{Ballance2013}.
In the work of Ballance and co-workers \cite{Ballance2013}, on  electron-impact excitation,
  over 10,000 close-coupling channels were included in the theoretical model, however 
in the current photoionization studies we are only approaching the 5000 channel mark. 

 In contrast to electron-impact excitation/ionization 
 calculations, photoionisation cross sections require only a few partial waves, governed by the dipole selection rules  
 as compared to typically 50-60 partial waves required by electron-impact collisions. 
 Furthermore, combined with currently available high resolution 
 experiments for certain ion stages of W, theoretical DARC photoionization cross section 
 calculations provide an ideal way to survey the first few ion stages of tungsten,
 indicating how well resonance structure is reproduced with a minimal atomic 9-12 
 configurations. This has future implications for the more intensive calculation of electron-ion recombination and ionization
 processes. In subsequent papers for the W$^{+}$-W$^{5+}$ ion stages, we intend to
 compare with more recent higher resolution experimental measurements already carried out at the 
 Advanced Light Source, in Berkeley. However, for the single photoionization of neutral 
 tungsten, only early experiments from the mid 1990s (which are not absolute) are available for comparison purposes. 
 These early  experiments  employed a time-resolved dual-laser plasma technique to measure the photo-absorption 
 of tungsten metal vapours \cite{Costello1991a}. These vapours were created by the ablation 
 of a spectroscopically pure tungsten solid, subject to a flashlamp pumped dye laser.
 A more complete description of the experiment is given in Costello \etal \cite{Costello1991a,Costello1991b}.
 Photo-absorption experiments of W in a solid, as well as other neighbouring elements in the periodic table (Ta,Re) were measured
 in the photon energy range 30-600 eV using synchrotron radiation  \cite{Haensel1969}, 
 at the 7.5-GeV electron synchrotron facility  DESY.  Sladeczek \etal \cite{sladeczek1995} 
 prepared the tungsten target by the thermal evaporation from the metal at a 
 temperature of 3200 K, to measure the photo-ion yield spectra of W$^+$ in the energy region of 30-60 eV.
 
 In the  current investigation we also show a theoretical comparison, for photoionization (PI) cross sections 
 between the present results obtained from a  Dirac Coulomb R-matrix approximation (DARC)  
 with the earlier Many-Bodied Perturbation Theory (MBPT) work of Boyle and co-workers \cite{Boyle1993} .

The remainder of this paper is structured as follows.
Section 2 presents a brief outline of the theoretical work. Section 3 presents a discussion of the
results obtained from both the experimental and theoretical methods.
Finally in section 4 conclusions are drawn from the present investigation. 
%
%
%
%
%
\section{Theory}\label{sec:Theory}
  
 An efficient parallel version \cite{ballance06}  of the DARC \cite{norrington87,norrington91,norrington04,darc} suite 
 of codes has been developed \cite{venesa2012,Ballance2012,McLaughlin2012}  to address the challenge of 
 electron and photon interactions with atomic systems catering for several thousands of scattering channels.
 Metastable states are populated in these experiments and therefore additional theoretical calculations  
 are carried out to gauge their extent.  
 Recent modifications to the Dirac-Atomic-R-matrix-Codes, as implemented for  
 (DARC)~\cite{venesa2012,McLaughlin2012,Ballance2012} allowed high quality photoionization cross section 
 calculations of Fe-peak elements and Mid-Z atoms of interest to astrophysics.  
 Cross-section calculations for trans-Fe element single photoionization of Se$^{+}$, Xe$^{+}$, Xe$^{7+}$, 
 Kr$^{+}$ ions \cite{Ballance2012,McLaughlin2012,Mueller2014b,Hino2012}, Si$^{+}$ ions \cite{Kennedy2014}, 
 and neutral Sulfur \cite{Barthel2015} have been made using the DARC code 
 and have shown suitable agreement with high resolution measurements made at third generation light sources.
 In particular for the calculations presented here, the capacity to assign an arbitrary number of processors 
 to the concurrent formation of every dipole matrix, mitigates the fact that matrix multiplications 
 of eigenvector matrices over 60,000 by 60,000 in size require at least ${\rm 2\times10^{14}}$ operations each.
 
 All the present photoionization cross section calculations were performed in 
 the Dirac Coulomb R-matrix approximation using our recently developed parallel version of the DARC 
 codes \cite{Ballance2012,McLaughlin2012,Mueller2014}.   
 The current state-of-the-art parallel DARC codes running on high performance computers (HPC)
world-wide, allows one to concurrently form and diagonalize 
large-scale Hamiltonian and dipole matrices \cite{McLaughlin2014a,McLaughlin2014b} required for
electron or photon collisions with atomic systems.
This enables large-scale cross-section calculations to be completed in a timely manner. 

\subsection{Tungsten (W)}

Photoionization cross sections on this complex system were performed for the ground and the excited metastable levels 
associated with the $\rm 5s^25p^65d^46s^2:$  $\rm ^5D$ term, and  
benchmarked with several available experimental  measurements.
We note that for neutral and or near neutral heavy atoms
determining their atomic structure sufficiently accurately is notoriously difficult.
The atomic structure calculations optimized on the residual W$^+$ ion were carried
out using the GRASP code \cite{dyall89,grant06,grant07}. 

Our first close-coupling calculation employed the lowest 645 energy accessible
 levels arising from the nine configurations of the W$^{+}$ residual ion: 
$\rm 4p^64d^{10}4f^{14}5s^25p^65d^46s$, 
$\rm 4p^64d^{10}4f^{14}5s^25p^65d^36s^2$,  
$\rm 4p^64d^{10}4f^{14}5s^25p^55d^46s^2$,
$\rm 4p^64d^{10}4f^{14}5s5p^65d^46s^2$, 
we also opened the $\rm 4f$-shell, namely,
 $\rm 4p^64d^{10}4f^{13}5s^25p^65d^46s^2$,    
$\rm 4p^64d^{10}4f^{13}5s^25p^65d^5$6s,
then the $\rm 4d$-shell,  $\rm 4p^64d^94f^{14}5s^25p^65d^46s^2$, 
the $\rm 4p$-shell, \\
 $\rm 4p^54d^{10}4f^{14}5s^25p^65d^46s^2$
 and finally the $\rm 4s$-shell, $\rm 4s4p^64d^{10}4f^{14}5s^25p^65d^46s^2$.
 
 In Table 1, we compare a representative sample of 
our energy levels with the NIST tabulated values  \cite{NIST2012}  for the residual ion W$^+$. 
We find that the maximum energy difference 
is typically of the order of $0.04$ Rydbergs for the valence ${\rm n=5 \ and \ n=6}$ shell levels, with 
no available energies given for the open $\rm n=4$ shell levels. 
This is well within the metastable uncertainty  of the experimental measurements.
 In the second calculation, we included all possible 1227 levels from the last 9 configurations in order 
 to complete the total photoionization of the remaining $\rm n=4$ shell up to a photon energy of 600 eV. The 
 4d direct ionization begins at 257.5 eV, the 4p direct ionization at approximately 442 eV and the 4s 
 direct ionization at 616 eV. 
 
\begin{table}
\begin{center}
\caption{Comparison of the theoretical energies using the GRASP code with the 
		NIST  \cite{NIST2012} tabulated data;  relative energies are in Rydbergs.
	       A sample of the 17 lowest NIST levels of the residual W$^{+}$ ion 
	       compared with our theoretical values from the 
	       645-level approximation are presented \label{structure}.}
\begin{indented}
\begin{tabular}{cccccc}
\br
Level  &  Configuration 		&  Term 							 & NIST  &GRASP 		& $\Delta^{\dagger}$   \\
		&				&			     					& Energy 			&Energy$^{\rm a}$ 	& 645		\\	
		&				&			     					& (Ryd)			&(Ryd) 			& levels 		 \\	
\mr
 1  		& $\rm 5d^4(^5D)6s$ 	&  $\rm ^6D_{\it 1/2}$          			& 0.000000		&0.000000		&   ~ 0.0    \\  
 2  		& $\rm 5d^4(^5D)6s$	&  $\rm ^6D_{\it 3/2}$          			& 0.013841		&0.009202		&     0.0046	\\
 3  		& $\rm 5d^4(^5D)6s$	&  $\rm ^6D_{\it 5/2}$          			& 0.028910		&0.021266		&     0.0076\\
 4  		& $\rm 5d^4(^5D)6s$	&  $\rm ^6D_{\it 7/2}$          			& 0.042978		&0.034552		&     0.0084	\\
 5  		& $\rm 5d^4(^5D)6s$	&  $\rm ^6D_{\it 9/2}$          			& 0.056016		&0.048546		&     0.0075	\\
 \\
6  		& $\rm 5d^5$			&  $\rm ^6S_{\it 5/2}$          			& 0.067618		&0.075682		&    0.0081   	\\
\\
7  		& $\rm 5d^36s^2$		&  $\rm ^4F_{\it 3/2}$          			& 0.079383		&0.119217		&    0.0398	\\
8  		& $\rm 5d^36s^2$		&  $\rm ^4F_{\it 1/2}$          			& 0.080490		&0.105601		&    0.0252     \\
9		& $\rm 5d^36s^2$		&  $\rm ^4F_{\it 3/2}$          			& 0.096526		&0.133134		&    0.0366	\\
10  		& $\rm 5d^36s^2$		&  $\rm ^4F_{\it 5/2}$          			& 0.102983		&0.142539		&    0.0396	\\
11  		& $\rm 5d^36s^2$		&  $\rm ^4F_{\it 7/2}$          			& 0.122219		&0.157617		&    0.0354	\\
\\
12  		& $\rm 5d^4(^5D)6s$ 	&  $\rm ^4D_{\it 1/2}$          			& 0.120044		&0.159874		&    0.0398	\\
13  		& $\rm 5d^4(^5D)6s$ 	&  $\rm ^4D_{\it 5/2}$          			& 0.122242		&0.150617		&    0.0284	\\
14  		& $\rm 5d^4(^5D)6s$ 	&  $\rm ^4D_{\it 3/2}$          			& 0.133358		&0.171186		&    0.0378	\\
15  		& $\rm 5d^4(^5D)6s$ 	&  $\rm ^4D_{\it 7/2}$          			& --				&0.173779		&    --	\\
16  		& $\rm5d^4(^5D)6s$ 	&  $\rm ^4D_{\it 9/2}$          			& 0.135388		&0.166293		&   0.0309 	\\
17  		& $\rm 5d^4(^5D)6s$ 	&  $\rm ^4D_{\it 5/2}$          			& 0.136396		&0.177468		&   0.0411	\\
\mr
\end{tabular}
\\
$^{\rm a}$ Theoretical energies from the 645-level approximation\\
$^{\dagger}\Delta$ Absolute difference in Rydbergs relative to NIST  tabulations \cite{NIST2012}.\\
\end{indented}
\end{center}
\end{table}

 \begin{table}
\begin{center}
\caption{Number of Scattering Channels associated with each dipole allowed symmetry}
\begin{tabular}{cccccc}
\br
Even Symmetry       &  Number of Channels &  Odd Symmetry & Number of Channels \\
\hline
${\it J}=0$               &         645         	&     ${\it J}^{\prime}=0$       &        645        \\
${\it J}=1$               &        1872         	&     ${\it J}^{\prime}=1$       &       1872        \\
${\it J}=2$               &        2929         	&     ${\it J}^{\prime}=2$       &       2929        \\
${\it J}=3$               &        3750         	&     ${\it J}^{\prime}=3$       &       3750        \\
${\it J}=4$               &        4320         	&     $\it J^{\prime}=4$       &       4320        \\
${\it J}=5$               &          -          		&     ${\it J}^{\prime}=5$       &       4671        \\
\mr
\end{tabular}
\\
\end{center}
\end{table} 
 Photoionization  cross section calculations for a 645-level and a 1227-level model were performed in 
 the Dirac-Coulomb approximation using the DARC codes \cite{Ballance2012,McLaughlin2012,McLaughlin2014a,McLaughlin2014b} 
 for the ground and metastable fine-structure levels associated with the  $\rm 5s^25p^65d^46s^2$  configuration 

 The R-matrix boundary radius of 12 Bohr radii  was sufficient to envelop
 the radial extent of the residual W$^{+}$ ion atomic orbitals. 
 A basis of 16 continuum orbitals was sufficient to span the incident experimental photon energy
 range from threshold up to 120 eV for the 645-level model and 30 continuum basis orbitals 
 to span 0-625 eV for the 1227 model. Since dipole selection rules apply, 
 total ground-state photoionization cross sections require  only the  
 ${\it  J}^{\pi}={\rm 0^{e}} {\rightarrow} {\it J} {^{\prime} {\pi^{\prime}}}={\rm 1^{\circ}}$, 
 bound-free dipole matrix.  In the case of the excited metastable states the 
${\it J}^{\pi}=4^{e} {\rightarrow} {\it J}^{{\prime} {\pi^{\prime}}}=5^{\circ},4^{\circ},3^{\circ}$,  
${\it J}^{\pi}=3^{e} {\rightarrow} {\it J}^{{\prime} {\pi^{\prime}}}=4^{\circ},3^{\circ},2^{\circ}$,
 ${\it J}^{\pi}=2^{e} {\rightarrow} {\it J}^{{\prime} {\pi^{\prime}}}=3^{\circ},2^{\circ},1^{\circ}$ and
  ${\it J} ^{\pi}=1^{e} {\rightarrow} {\it J}^{{\prime} {\pi^{\prime}}}=2^{\circ},1^{\circ},0^{\circ}$ are necessary.
 
We note that the present investigations are currently the largest photoionization calculations
in terms of the number of scattering channel performed to date with our DARC codes. 
For example, in the case of the initial ${\it J}=4$ metastable level, we have a ${\it J}=5^o$ symmetry with 4,761 scattering channels 
and matrices of the size of 60,723 to consider in the photoionization calculation. 
In table 2, we list the number of channels associated with each 
of the ${\it J} - {\it J}^\prime$ dipole pairs used in the statistically averaged comparison 
with experiment, also allowing for the fact that the $\rm 5d^56s, \ {\it J}=3$ excited level 
lies energetically below the $\rm  5d^46s^4, \ {\it J}=3$ of the neutral tungsten ground state configuration.

\subsection{Photoionization cross sections}

 For the ground and metastable initial states of the neutral W atom studied here, 
 the outer region collision problem was solved (in the resonance region below and
 between all thresholds) using a suitably chosen fine
energy mesh of 6.0$\times$10$^{-4}$ Rydbergs ($\approx$ 8.0 meV), in order 
to resolve all important resonance structure in the appropriate photoionization cross sections. 
This is sufficient for comparison purposes with the previous experimental 
results \cite{Haensel1969,Costello1991a,sladeczek1995} which were taken at resolutions
ranging from 0.25 eV (250 meV) to 0.5 eV (500 meV).
 
To simulate the experimental measurements, the DARC theoretical photoionization cross section results have been convoluted with a Gaussian 
having a profile of full width half maximum (FWHM) similar to the experimental resolution.
In order to compare with experiment, we have statistically weighted the ground and metastable photoionization
cross sections as well as the initial bound level.
This has the effect of reducing the actual photoionization threshold, due to the excited states of the $\rm ^5D$ having consistently 
higher quantum numbers. Comparatively, it also reduces the absolute heights of the resonance structure between 40 
and 60 eV as opposed to the ${\it  J}=0 - 1$ calculation, but otherwise the individual 5 level-resolved cross sections are remarkably similar.

We note that there are other metastable levels from the same initial $\rm 5d^46s^2$ configuration up to at least 0.285 Rydbergs 
above the ground state, leading to energy differences of up to 3 eV when comparing the energy scale
with experimental measurements. A complete direct modelling  comparison with experiment 
would required the calculation of photoionization cross sections from all the 34 levels of the ground state 
configuration, and probably approximately 50 levels from the $\rm 5d^56s$ and 
$\rm 5d^46s6p$ configurations that are interspersed among them. However, as we 
only have access to finite computational resources 
these extra 79 calculations in addition to the 5 presented here
 are beyond present capabilities. 

\section{Results and Discussion}

 The calculated ionization potential from our DARC calculation is 8.09 eV which is 
 230 meV above the NIST tabulated value of 7.86 eV. However the statistically averaged 
 initial bound state reduces this difference to a value of 170 meV, bringing it closer to the experimental value.

\begin{figure}
\begin{center}
\includegraphics[scale=1.25,width=\textwidth]{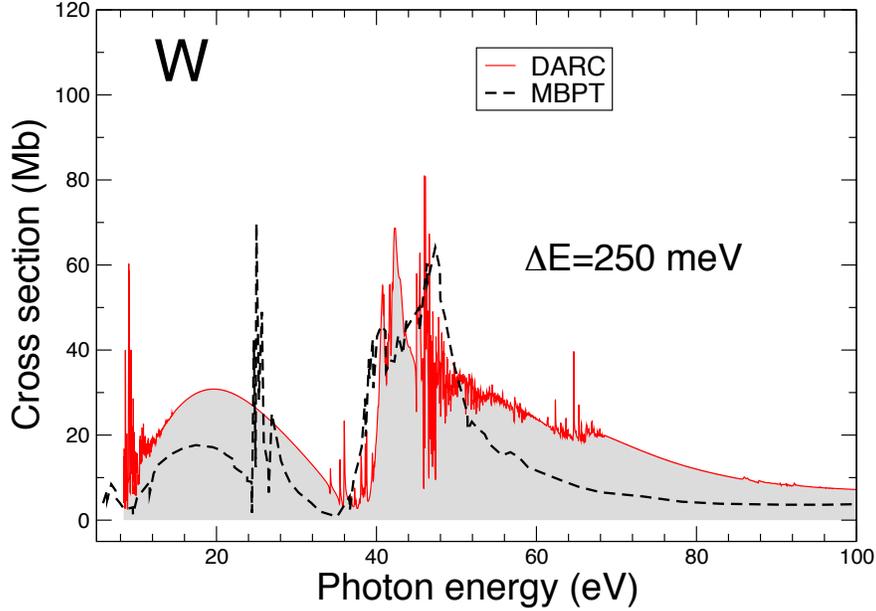}
\caption{\label{Wfig1}(Colour online) Single photoionization of neutral W over the 
				photon energy range 8 eV - 100 eV. The solid line and corresponding 
				shaded area correspond to the total photoionization 
				of the 6s, 5p, 5s and 4f orbitals from the ${\it J} =0 : 6s^2$ tungsten ground state.   
				Cross sections are given in Mb. The present DARC calculations were
				Gaussian convolved at FWHM 250 meV. The dashed line 
				corresponds to the earlier MBPT \cite{Boyle1993} calculation for the ground state level.}
\end{center}
\end{figure}

 As Fig.\ref{Wfig1} illustrates, collectively the direct photoionization of the 6s and 5d orbitals 
 peak at 30 Mb at a photon energy of 19 eV. However, it is the direct photoionization of the
 4f and the 5p orbitals, and the associated ${\rm 4f \ - \ 5d}$ and ${\rm 5p \ - \ 5d}$ resonances that dominate the cross 
 section from 40-65 eV, with narrow peaks as high as 75-80 Mb. The 5s photoionization 
 is evident from 65 eV onwards, but only marginally so. The onset of our 4f photoionization 
 at approximately 38-39 eV slightly leads the 5p photoionization by 1-2 eV, and it 
 agrees well with the MBPT theory in terms of energy position. 
 However the MBPT theoretical cross sections are persistently lower across the entire energy range.
 Between 43 eV and 54 eV, our model has several hundred thresholds belonging 
 to either $\rm 4p^64d^{10}4f^{13}5s^25p^65d^46s^2$ or $\rm 4p^64d^{10}4f^{14}5s^25p^55d^46s^2$    
 configurations thus making it difficult to determine the relative strength of each.
 The sharp resonance at 35 eV, identified in the Sladeczek experiment
 \cite{sladeczek1995}  which they attributed to a ${\rm 5p - 6s}$ transition,
 is well reproduced by the current DARC model. 

\begin{figure}
\begin{center}
\includegraphics[scale=1.25,width=\textwidth]{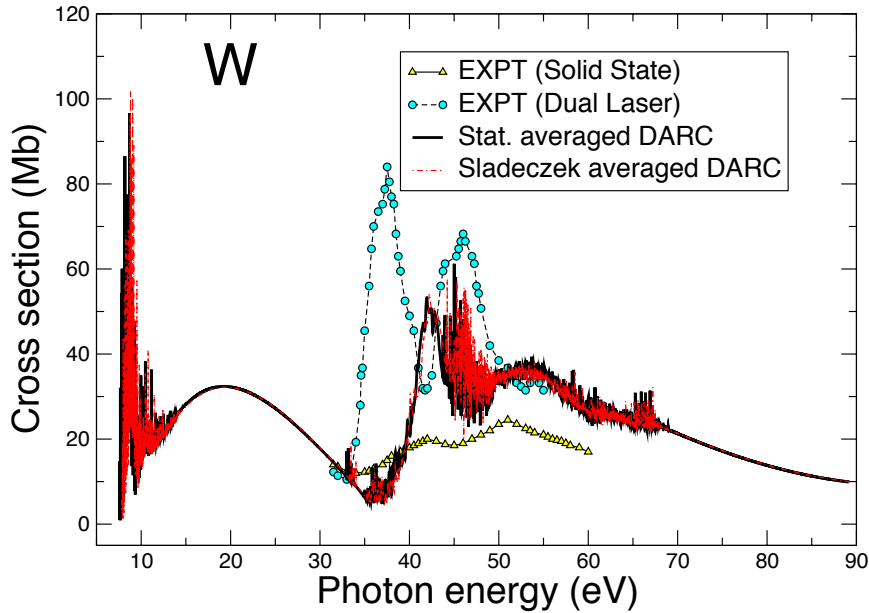}
\caption{\label{Wfig2}(Colour online) Single photoionization of neutral W over the 
					photon energy range 8 eV - 100 eV, comparing weighted averaged theoretical 
					calculations with several different experiments. The DARC results (645-levels approximation, solid black line), 
					are the statistical average of the 5 levels associated the $\rm ^5D$ term ground state. 
					The (dashed  red-line) is the weighted DARC results of the lowest 6 levels, employing the mixing coefficients reported 
					by Sladeczek and co-workers \cite{sladeczek1995}.  Solid circles with dashed line are the dual-laser experimental 
					results of Costello and co-workers \cite{Costello1991a}  and the solid triangles are the results
					from the experimental work of Haensel and co-workers \cite{Haensel1969}. 
					The statistically averaged DARC PI cross sections were Gaussian convolved at a FWHM of  250 meV.} 
\end{center}
\end{figure}

 In Fig.\ref{Wfig2}, we compare the ground state term statistically averaged PI cross section results with the experiments
 of Costello \etal \cite{Costello1991a} and those of Haensel \etal \cite{Haensel1969}. 
 At 35 eV, the dual laser experiment exhibits the  onset of the 4f ionization at approximately 3 - 4 eV, 
 before both the theoretical values of the MBPT \cite{Boyle1993} 
 and the present DARC PI cross section calculations, 
 however still within the range of the other higher levels of 
 the $\rm 6s^2$ configuration as reported in the NIST energy level table. 
 This suggests that the metastable component of the experiment of Costello and 
 co-workers \cite{Costello1991a} may include some of these higher levels. 
 The Haensel experiment \cite{Haensel1969}, from a solid-state target does not exhibit the strong 4f-5d, or 5p-5d
 resonance structure to the same extent compared to all presented
 theories and the dual-laser experiment. The Sladeczek experiment \cite{sladeczek1995} 
 (also from a solid perspective), does not provide absolute values either, but in terms of 
 a relative minimum to maximum ratio of the measurement is comparable 
 to the Haensel and co-workers \cite{Haensel1969}  experimental result.   
 Sladeczek and co-workers \cite{sladeczek1995} do 
 however  report a relativistic Hartree-Fock calculation which is a mixture 
 of the first six levels of neutral tungsten. We have employed these same six 
 mixing coefficients with our present DARC PI cross section results, represented by the 
 dashed-line in Fig.\ref{Wfig2}.  Not surprisingly, as the individual PI
 cross sections are remarkably similar, (shifted slightly only by energy differences in the target), 
 it provides a result very close to the prediction of the statistically averaged DARC result.  

\begin{figure}
\begin{center}
\includegraphics[scale=1.25,width=\textwidth]{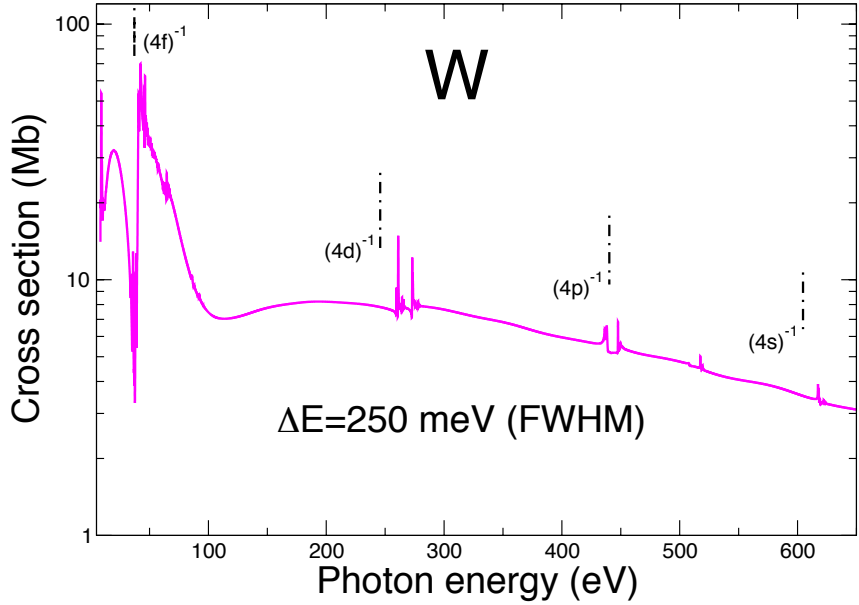}
\caption{\label{Wfig3}(Colour online) Single photoionization cross sections of neutral W over the photon energy range 8 eV - 625 eV,
				from the present DARC PI calculations incorporating 1227 levels in the close-coupling calculations. 
				The photoionization threshold of each of the $4\ell^{-1}$ orbitals hole configurations are 
				indicated on the graph.}
\end{center}
\end{figure}

Comparison with other theoretical methods is a much simpler prospect than with experimental measurement. 
There are no unwanted metastable components to contend with 
and it requires the calculation of a single ${\rm {\it J}=0-{\it J}{^\prime}=1}$ dipole matrix as is  illustrated in Fig.\ref{Wfig1}. 
The reduced number of associated channels with partial waves of low angular momentum 
has allowed us to increase the photon energy range up to 625 eV, 
as illustrated in Fig.\ref{Wfig3}. Here we extend our 
DARC PI cross section calculations to span the energy to encompass 
 the photoionization energy range of all the {\it n}=4, 5 and 6 shells. Above 100 eV, 
 the direct ionization of the 4d, 4p and 4s orbitals is minimal compared with
 the preceding sub-shells. The Rydberg resonances attached to the upper levels 
 of each hole configuration are much more evident than the direct total photoionization 
 above 200 eV. The photoionisation of the 4f orbital is the dominant 
 component of the total photoionisation cross section in this energy range.
	%
	%
	%
	%
	%
	%

\section{Conclusions}\label{sec:Conclusions}

Photoionization cross sections calculations for the neutral tungsten atom were 
obtained from large-scale close-coupling calculations within
the Dirac-Coulomb R-matrix approximation (DARC) 
and compared with moderately resolved experimental measurements.

The neutral tungsten photoionization presented here shall complement a future series of studies
on tungsten ions W$^{q+}$ in low charged states up to $q=5$ \cite{Mueller2011,Mueller2012,Mueller2014}. 
The comparison of the measured photoionization spectra with large-scale R-matrix close-coupling calculations
presented here shows good agreement, and it is expected that 
applying a similar approach to other charged states of tungsten that
have been investigated by more recent experiments at ALS 
will provide valuable intepretation of those measurements. 
It is hoped that the present theoretical work will 
act as a stimulus and encourage new experimental studies on neutral tungsten 
photoionization at the same high resolution achieved by the other ion stages of this system.

\ack
C P Ballance was supported by NSF and NASA grants  through Auburn University.
B M McLaughlin acknowledges support by the US National Science Foundation through a grant to ITAMP
at the Harvard-Smithsonian Center for Astrophysics, Queen's University Belfast 
for the award of a visiting research fellowship (VRF) and the hospitality 
of the University of Auburn during a research recent visit.
The computational work was carried out at the National Energy Research Scientific
Computing Center in Oakland, CA, USA and at the High Performance 
Computing Center Stuttgart (HLRS) of the University of Stuttgart, Stuttgart, Germany. 
This research also used resources of the Oak Ridge Leadership Computing Facility 
at the Oak Ridge National Laboratory, which is supported by the Office of Science 
of the U.S. Department of Energy under Contract No. DE-AC05-00OR22725.
Helpful discussions with Professor E T Kennedy are gratefully acknowledged.
\clearpage
%
%
%
%
\section*{References}
\bibliographystyle{iopart-num}
\bibliography{wions}

\end{document}